\documentclass[prx,twocolumn,groupedaddress]{revtex4}
\usepackage[utf8]{inputenc}

\usepackage{graphicx}
\usepackage{late xsym}
\usepackage{amssymb}
\usepackage{amsmath}
\usepackage{amsfonts}
\usepackage{gensymb}
\usepackage{bm}
\usepackage{braket}
\usepackage{physics}
\usepackage{bbold}
\usepackage{cases}
\usepackage{tabularx}
\usepackage{multirow}
\usepackage{braket}
\usepackage{epstopdf}
\usepackage[urlcolor=blue]{hyperref}

\newcommand{\be}{\begin{equation}}
\newcommand{\ee}{\end{equation}}

\newcommand{\bea}{\begin{eqnarray}}
\newcommand{\eea}{\end{eqnarray}}
\usepackage{color}
\usepackage{dcolumn}
\usepackage{epsfig}
\usepackage{bm}
\usepackage[urlcolor=blue]{hyperref}
\hypersetup{backref, colorlinks=true, linkcolor=blue, citecolor=blue}

\begin{document}

\title{Orbital Fulde-Ferrell-Larkin-Ovchinnikov state in an Ising superconductor}

\author{Noah F. Q. Yuan}
\email{fyuanaa@connect.ust.hk}
\affiliation{School of Science, Harbin Institute of Technology, Shenzhen 518055, China}

\begin{abstract}
The critical field behavior of a layered Ising superconductor with finite number of layers is studied. Under in-plane magnetic fields, the finite-momentum superconductivity dubbed as the orbital Fulde-Ferrell-Larkin-Ovchinnikov state is found in the regime of low field and high temperature.
Our theory agrees well with the experimental results in Nature \textbf{619}, 46 (2023).
\end{abstract}

\maketitle

\textit{Introduction}---
The dimensionality of a superconductor is usually derived from its critical field behavior, which reflects the spatial profile of the order parameter under external magnetic fields.
When the superconductor is three-dimensional (3D), Abrikosov vortices \cite{Abrikosov} will be formed under fields above the lower critical field, where the magnitude of the order parameter is highly non-uniform and the phase has an integer winding around a vortex core. As a result, the upper critical field of a 3D superconductor is linear in temperature, and the critical exponent is independent of the field direction.

On the contrary, in a two-dimensional (2D) superconductor the order parameter is mostly uniform in magnitude and Abrikosov vortices cannot be formed under in-plane fields. Since the characteristic size of the Abrikosov vortex core is the coherence length $\xi$, the criterion of 2D superconductivity is roughly $d<\xi$, where $d$ is the thickness. Detailed calculations further reveal the critical thickness for 2D superconductivity $d<d_c\approx 1.8\xi$ \cite{tinkham}. In a 2D superconductor, near the zero-field critical temperature, the out-of-plane upper critical field is still linear in temperature, while the in-plane one is square-root in temperature \cite{tinkham,tinkhamPR}, as verified in experiments \cite{Harper,JMLu,XXi,YuS}. 

The above arguments on dimensionality are based on continuum models of superconductors. Over the past several decades, the layered superconductors have been discovered \cite{Bul,klem,tinkham,YuS}, where each layer is an atomically thin 2D superconductor, and different layers are weakly coupled by Josephson coupling \cite{J1,J2,Lik,Klei1,Klei2}. 
For a layered superconductor with infinite number of layers, a dimensional crossover can be realized by an in-plane magnetic field \cite{tinkham,klem,LD,KLB,KBL}. When the in-plane field is weak compared with the interlayer Josephson coupling, the layered superconductor can be treated as 3D with an upper critical field linear in temperature. As the field increases so that the interlayer coupling is relatively negligible, the layered superconductor behaves as if a 2D superconductor with an upturn in the upper critical field.
Such a dimensional crossover has been experimentally observed in several layered superconductors \cite{Rugg,Prob,KLB,KBL}.

Recently, it was experimentally found \cite{nature} that a layered Ising superconductor NbSe$_2$ with intermediate thickness can have unconventional critical field behavior. The upper critical field is square-root in temperature and hence 2D at low fields. As the field increases, an additional phase transition associated with a tricritical point is found instead of a dimensional crossover from 3D to 2D. 
These results are inconsistent with the critical field behavior of the layered superconductor with infinite layers \cite{tinkham,KBL,KLB,LD}, but show similarities to bilayer superconductors \cite{CXL,GWQ} instead. In Refs. \cite{CXL,GWQ}, it is found that in a bilayer superconductor linked by Josephson coupling, unconventional superconducting states similar to the Fulde-Ferrell-Larkin-Ovchinnikov (FFLO) states \cite{FF,LO} can be realized under in-plane magnetic fields mainly through the orbital effect.

\begin{figure}
\includegraphics[width=\columnwidth]{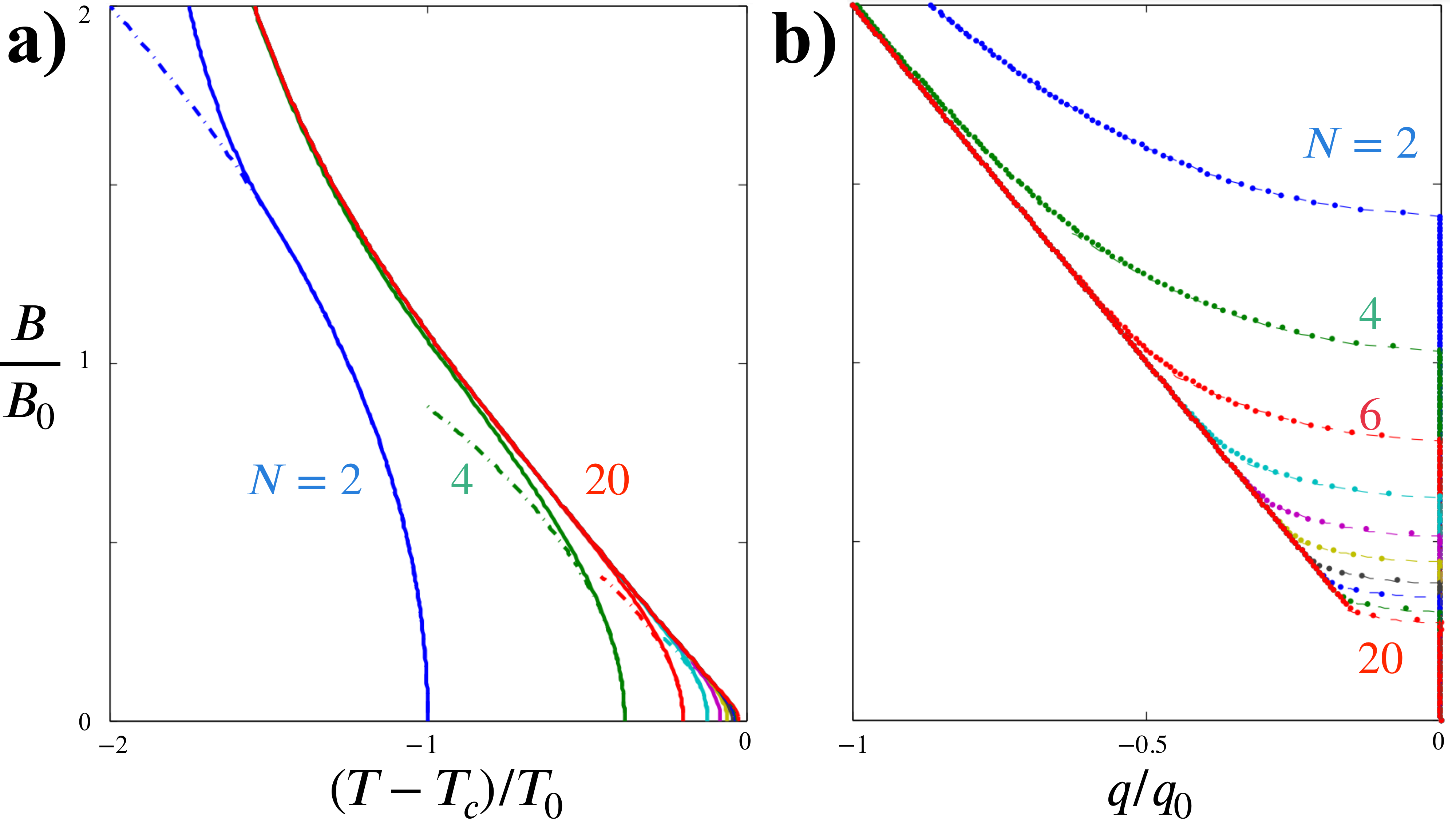}
\caption{Orbital FFLO phases in $N$-layer superconductors. a) Upper critical field as a function of temperature. With the increasing zero-field critical temperature, even number $N$ changes from 2 to 20, labeled by different colors. Solid lines are numerical results, dashed lines denote the analytical formula Eq. (\ref{eq_tcnb}).
b) Upper critical field versus Cooper pair momentum. Colors denote $N$, the same as a). Dashed lines denote the analytical formula Eq. (\ref{eq_q}). 
As $N$ increases, upper critical field approaches the envelope $B=B_0(T_{c}-T)/T_0$ in a) and optimal Cooper pair momentum approaches the envelope $q/q_0=-\frac{1}{2}B/B_0$ in b). Here $B_0,T_0,q_0$ are defined in Eq. (\ref{eq_units}).}\label{fig1}
\end{figure}

In this manuscript, we would like to study the layered superconductors with intermediate number of layers under in-plane fields, and try to generalize the FFLO-like states in Refs. \cite{CXL,GWQ} to multilayer cases. 

It is well known that two depairing mechanisms are introduced by an external magnetic field, namely the orbital effect and the Zeeman effect. 
In a conventional superconductor, the Zeeman effect can be neglected at weak fields far below the Pauli limit $B_{P}\equiv\Delta_{0}/\sqrt{2}\mu$ \cite{WHH,KLB,KBL}, where $\Delta_{0}$ is the pairing gap at zero temperature and $\mu$ is the magnetic moment.
In an Ising superconductor, the Zeeman effect is screened \cite{JMLu,XXi,YuS} for in-plane fields far below the Ising limit $B_{so}\equiv\Delta_{so}/\mu$, where $\Delta_{so}$ is the Ising spin-orbit coupling (SOC) energy.
As the materials we are interested in (e.g. Ising superconductors of transition metal dichalcogendies) are composed of bilayer unit cells, in the following we consider only even number of layers if not specified otherwise.

\textit{Model}---We consider the $N$-layer Lawrence-Doniach (LD) model with the following free energy density per area \cite{LD,GWQ,CXL}
\bea\label{eq_LD}
f&=&\sum_{l=1}^{N}\left\{\alpha|\psi_l|^2+\frac{\beta}{2}|\psi_l|^4+\frac{|(\nabla_{\parallel}-2ie\bm A_l)\psi_l|^2}{2m}\right\}\\\nonumber
&-&J\sum_{l=1}^{N-1}(\psi_{l}^{*}\psi_{l+1}+\psi_{l+1}^{*}\psi_{l})\exp\left(2ie\int_{ld}^{(l+1)d} A_z dz\right),
\eea
where $m>0$ is the electron mass, $e$ the electron charge, and $J>0$ is the Josephson coupling between nearest neighbor layers. In the gradient terms, $\nabla_{\parallel}=(\partial_x,\partial_y)$ is the in-plane gradient operator, $\bm A_l$ is the in-plane vector potential on layer $l$, and $A_z$ is the out-of-plane component of vector potential.
The second order coefficient can be worked out as a function of temperature $T$ and field $B$
\be\label{eq_mono}
\alpha =N_0\left\{
\log\left(\frac{T}{T_{c1}}\right)-\frac{B^2}{B^2+B_{so}^2}F\left(\frac{\mu\sqrt{B^2+B_{so}^2}}{\pi T}\right)
\right\},
\ee
where $N_0$ is the density of states of each layer, the special function $F(x)={\rm Re}\left\{\Psi(\frac{1}{2})-\Psi\left[\frac{1}{2}(1+ix)\right]\right\}$ is defined in terms of the digamma function $\Psi(x)$, $T_{c1}$ is the critical temperature of the monolayer. The fourth order coefficient $\beta>0$ can be treated as a positive constant independent of temperature and field.
When $B\ll B_{so}$, the field-dependent term in Eq. (\ref{eq_mono}) can be neglected, and we can write $\alpha=\alpha_0(T-T_{c1})$ when $T\lesssim T_{c1}$, where $\alpha_0=N_0/T_{c1}$. In other words, the Zeeman effect of in-plane fields far below Ising limit can be neglected in a layered Ising superconductor.



Now we consider finite number of layers.
We employ the 2D ansatz $\psi_l=\Delta_l e^{i\bm q\cdot\bm r}$, whose magnitude is in-plane uniform $|\psi_l(\bm r)|\equiv\Delta_{l}$, and the phase is characterized by Cooper pair momentum $\bm q$. 
Correspondingly we choose the gauge $A_z=0$, $\bm A_l=[l-\frac{1}{2}(N+1)]d\bm B\times\hat{\bm z}$.
Hence the free energy is $f=f(\{\Delta_l\},\bm q,\bm B)$. 
Then the upper critical field of the second order superconducting phase transition is given by the highest critical field among all $\bm q$ and order parameter profile $\{\Delta_l\}$, and the corresponding $\bm q$ is the optimal Cooper pair momentum.

In the following, we mainly focus on the upper critical field regime, hence $B$ usually denotes the in-plane upper critical field if not specified otherwise.

\begin{figure}
\includegraphics[width=\columnwidth]{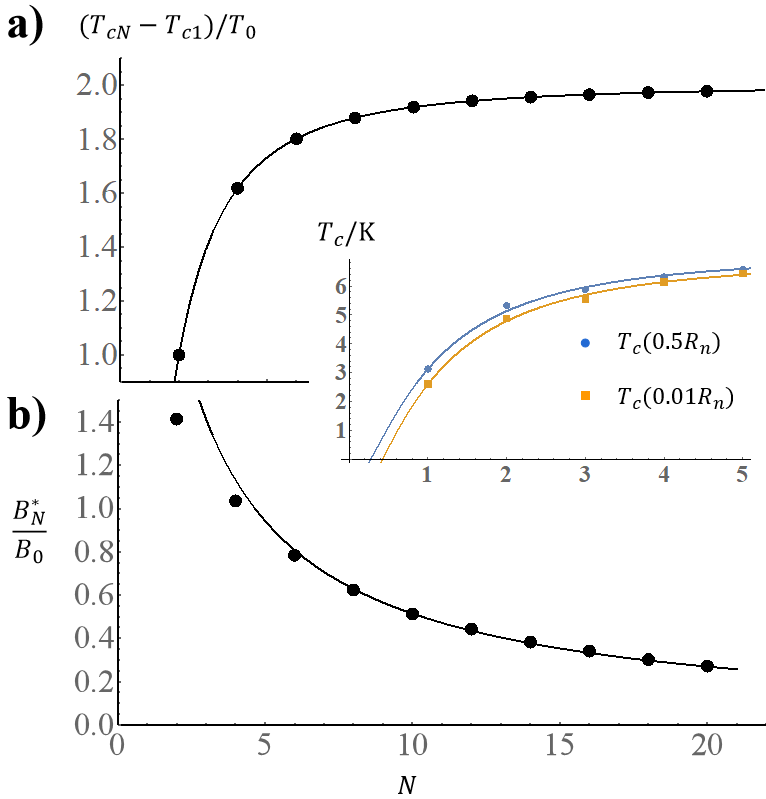}
\caption{a) Zero-field critical temperature $T_{cN}$ as a function of layer number $N$. Dots are from numerical simulations as in Fig. \ref{fig1}a and the black line is Eq. (\ref{eq_TcN}). Inset: Fittings (lines) of experimental data (markers) in Ref. \cite{XXi} by Eq. (\ref{eq_TcN}). Different colors denote $T_c$ values according to different criteria, and $R_n$ is the normal resistance.
b) Tricritical field $B_N^*$ as a function of layer number $N$. Dots are from numerical simulations as in Fig. \ref{fig1}b and the black line is Eq. (\ref{eq_BN}).}\label{fig2}
\end{figure}

\textit{Results}---The upper critical field and optimal Cooper pair momentum are numerically calculated and plotted in Fig. \ref{fig1}a and b respectively, normalized by the three units of temperature, field and momentum respectively
\be\label{eq_units}
T_0= \frac{J}{\alpha_0},\quad B_0=\frac{\Phi_0 q_0}{2\pi d},\quad q_0=\sqrt{2mJ},
\ee
where $\Phi_0=h/(2e)$ is the flux quantum.

At zero field, superconductivity occurs at temperatures below the layer-dependent critical temperature $T_{cN}$ as plotted in Fig. \ref{fig2}a. 
Similar results can be found in Ref. \cite{Sch}, while the mechanism is interlayer Cooper pairs instead of interlayer Josephson coupling. The zero-field critical temperature data of few-layer NbSe$_2$ can be found in Ref. \cite{XXi}, which agree well with Eq. (\ref{eq_TcN}) as shown in the inset of Fig. \ref{fig2}a.
This result can be analytically obtained as shown in the next session.

As shown in Fig. \ref{fig1}b,
at low fields, the Cooper pair momentum remains zero, and the layered superconductor 
behaves as a 2D superconductor with square-root temperature-dependence of the upper critical field as shown in Fig. \ref{fig1}a (dahsed lines). 
When the field further increases, the optimal Cooper pair momentum becomes finite and along $\pm\bm B\times\hat{\bm z}$ directions, which applies to all even numbers of $N\geq 2$. 
We define $q\equiv\bm q\cdot(\hat{\bm B}\times\hat{\bm z})$, which is plotted in Fig. \ref{fig1}b as a function of field strength $B$, implying a phase transition with a tricritical point $(T_{N}^*,B_N^*)$.
When $B<B_{N}^*$, the Cooper pair momentum is zero $q=0$, When $B>B_{N}^*$, $q\neq 0$, and at the tricritical point $(T^*,B^*)$, two types of superconducting phases $q=0$ and $q\neq 0$ coexist with the normal phase. The numerical tricritical field $B_N^*$ is plotted for different $N$ in Fig. \ref{fig2}b. 
In our finite-layer model, inversion symmetry is found to hold for the free energy, under which layer $l$ with momentum $\bm q$ maps to layer $N+1-l$ with momentum $-\bm q$. Thus states with $\pm\bm q$ are degenerate in calculating the upper critical field, and in Fig. \ref{fig1}b we only plot the non-positive branch of $q$.

The superconducting phase with finite-momentum Cooper pairs is not rare at least in theoretical studies. In 1964, Fulde and Ferrell \cite{FF}, Larkin and Ovchinnikov \cite{LO} proposed that due to Zeeman effect, magnetic field can drive Cooper pairs into finite-momentum states. Since the energy saved by finite-momentum Cooper pairs per unit field is small, conventional FFLO states are expected at low temperatures and high fields, which turn out not easy to be detected experimentally. However, in our theory, the finite-momentum Cooper pairs are boosted by magnetic field via orbital effect, which can survive at relatively low field and high temperatures. We may dub such states as the orbital FFLO states.


\begin{figure}
\includegraphics[width=\columnwidth]{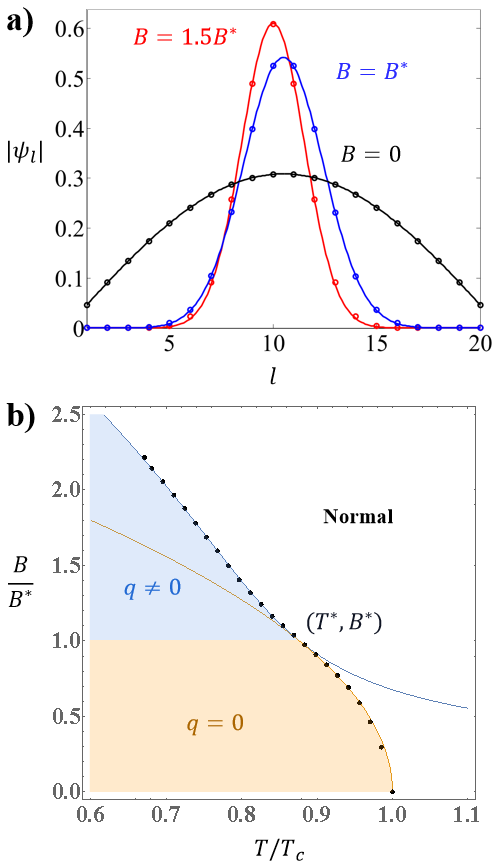}
\caption{a) Order parameters at different in-plane fields with layer number $N=20$. Open circles are numerical results of Eq. (\ref{eq_LD}) and solid lines are analytical formulae Eqs. (\ref{eq_psi0},\ref{eq_psiB}). b) Phase diagram of the effective model Eq. (\ref{eq_A}). Black dots are experimental data of upper critical field in Ref. \cite{nature}, while orange and blue lines are analytical results. BCS state $q=0$, orbital FFLO state $q\neq 0$ and normal phase coexist at the tricritical point $(T^*,B^*)$.}\label{fig3}
\end{figure}

\textit{Analysis}---To figure out the origin and mechanism of the orbital FFLO states, we analyze the order parameter profiles near superconducting phase transitions. 

We first review the inifinite layer limit $N\to\infty$. In this limit, due to the translation symmetry $l\to l+1$, the order parameter at zero field is uniform, and the bulk critical temperature is $T_{cN}\to T_c\equiv T_{c1}+2T_0$.
At finite in-plane fields, Abrikosov vortices can be formed, and the bulk in-plane upper critical field is $B_{c2}=B_0(T_c-T)/T_0$.

When $N$ is finite, the translation symmetry is lost, and the middle layers become more favorable than other layers because of the finite size effect. Under an in-plane field, Abrikosov vortices become metastable, and Josephson vortices formed by middle layers are the stable state, which are the orbital FFLO states we found previously.

To be concrete, the order parameter at zero field reads
\be\label{eq_psi0}
\psi_l\propto \sin\left(\frac{\pi l}{N+1}\right),\quad (B=0)
\ee
which is nonuniform, and the middle two layers $l=\frac{1}{2}N$ and $l=\frac{1}{2}N+1$ have the highest weightings.
From Eq. (\ref{eq_psi0}), the critical temperatures at zero and weak fields are
\bea\label{eq_TcN}
T_{cN}=T_{c1}+2T_0\cos\left(\frac{\pi}{N+1}\right),\quad (B=0)\\\label{eq_tcnb}
T_{cN}(B)=T_{cN}-c_N T_0\left(\frac{B}{B_0}\right)^2.\quad (B\ll B_0)
\eea
Detailed derivations of the order parameter Eq. (\ref{eq_psi0}), the critical temperatures Eqs. (\ref{eq_TcN},\ref{eq_tcnb}), and the expression of the dimensionless coefficient $c_N$ can be found in the Appendix.
Importantly, the quadratic dependence of the field in Eq. (\ref{eq_tcnb}) implies the 2D characteristic of the in-plane upper critical field in the weak field regime, as shown in the dashed lines of Fig. \ref{fig1}a and found experimentally in Ref. \cite{nature}.

As shown in Fig. \ref{fig3}a, at zero field, the order parameter profile Eq. (\ref{eq_psi0}) is peaked in the middle between layer $l=\frac{1}{2}N$ and $l=\frac{1}{2}N+1$, with a broad width. 
At weak field, the order parameter is still peaked at in the middle between layer $l=\frac{1}{2}N$ and $l=\frac{1}{2}N+1$, but with a smaller width. As field increases, the width of order parameter keeps shrinking. 
When the field is strong enough, the width of order parameter becomes small enough, Cooper pairs tend to accumulate on layer $l=\frac{1}{2}N$ or $l=\frac{1}{2}N+1$.
To be more precise, the order parameter profile at intermediate in-plane fields is a Gaussian profile \cite{tinkham}
\bea\label{eq_psiB}
\psi_l\propto \exp\left(-\frac{1}{2}\frac{|l-l_0|^2}{l_B^2}\right),
\eea
with field-dependent width $l_B=\sqrt{B_0/B}$ and center
\bea
l_0=
\begin{cases}
\frac{1}{2}(N+1) & (B\lesssim B^*_N)\\
\frac{1}{2}N\ {\rm or}\ \frac{1}{2}N+1 & (B> B^*_N)
\end{cases}.
\eea

At zero field, the order parameter can be characterized by the typical width $\delta l_N\sim \sqrt{N+1}/\pi$ (see Appendix). 
The order parameter becomes sufficiently localized when $l_B\lesssim\delta l_N$, or equivalently when $B\gtrsim B^*_N$, with the tricritical field
\be\label{eq_BN}
B_{N}^{*}\approx \frac{1.6\pi B_0}{N+1},
\ee
and tricritical temperature $T_{N}^{*}\approx T_{c}-{1.6\pi T_0}/{(N+1)}.$
 


The critical behavior near the tricritical point $(T^*,B^*)$ is governed by an effective bilayer model of two localized modes $\psi_{\pm}$ on layer $\frac{1}{2}N$ and $\frac{1}{2}N+1$ respectively. 
One then expands the free energy density as
$f=\Psi^{\dagger}\mathcal{H}\Psi+\dots$ on the basis $\Psi=(\psi_{+},\psi_{-})^{\rm T}$, and up to $q^2$ and $B^2$:
\bea\label{eq_A}
    \mathcal{H}_0(\bm q,\bm B)=a+b(\hat{\bm z}\times\bm B)\cdot\bm q\tau_z+cq^2-\mathcal{J}\tau_x,
\eea
where $\bm\tau$ denote Pauli matrices acting on the $\Psi$ basis,  $a=r(T-T_a)+\chi B^2$, and $r,\chi,b,c,\mathcal{J},T_a$ are phenomenonlogical parameters. For the stability of the free energy, we require $r,\chi,c$ to be positive.
The $\bm q,\bm B$ coupling term is due to the orbital effect, while the magnetic energy term $\chi B^2$ can be induced by both orbital and Zeeman effects of the in-plane magnetic field. 

At zero field, the critical temperature of the effective bilayer model Eq. (\ref{eq_A}) is $T_c=T_a+\mathcal{J}/r$.
Under an in-plane field, the optimal Cooper pair momentum is
\be\label{eq_q}
\bm q=\frac{b}{2c}(\bm B\times\hat{\bm z}){\rm Re}\sqrt{1-\left(\frac{B^{*}}{B}\right)^4}
\ee
with the tricritical field $B^*={\sqrt{2c|\mathcal{J}|}}/{b}$. 


The field-dependence of optimal Cooper pair momentum is plotted in Fig. \ref{fig1}b (dashed lines), which agrees well with numerical simulations (dots).
Correspondingly, the upper critical field as a function of temperature is plotted in Fig. \ref{fig3}b (orange and blue lines), which shows a kink at the tricritical point, and agrees well with experimental data from Ref. \cite{nature} (black dots).
Details of the fitting parameters can be found in Appendix.

Symmetry principles also allow us to write down higher order terms. In transition metal dichalcogenide such as NbSe$_2$, the point group is $D_{3d}$, and the sixth order terms are
$\mathcal{H}_1(\bm q,\bm B)=\lambda_1{\rm Re}(q_{+}^2B_{+}^4)+\lambda_2{\rm Re}(q_{+}^4B_{+}^2)+\lambda_3{\rm Im}(q_{+}B^2_{+})\tau_y$,
where $q_{\pm}=q_x\pm iq_y$, $B_{\pm}=B_x\pm iB_y$, and the total Hessian matrix is $\mathcal{H}=\mathcal{H}_0+\mathcal{H}_1$. These higher order terms leads to the anisotropy of the field-dependent critical temperature in the orbital FFLO phase with $q\neq 0$, while the field-dependent critical temperature in the conventional superconducting phase is isotropic since $q=0$. To be concrete, by plugging Eq. (\ref{eq_q}) into $\mathcal{H}_1$ we can calculate the anisotropic part of the field-dependent critical temperature
$\Delta T_c(B,\varphi)=\lambda(B)\theta(B-B^*)\cos(6\varphi),$
where $\bm B=B(\cos\varphi,\sin\varphi)$, $\theta$ is the Heaviside theta function, and $\lambda(B)$ is given in Appendix. 

The leading anisotropy is sixfold due to the following symmetry reasons.
Since there is no intrinsic time-reversal symmetry breaking in our model, $T_c(-\bm B)=T_c(\bm B)$. The threefold in-plane rotation symmetry $C_{3z}$ of TMD implies $T_c(C_{3z}\bm B)=T_c(\bm B)$, we derive that the leading anisotropy of the orbital FFLO states in TMD is sixfold 
$T_c(C_{6z}\bm B)=T_c(\bm B)$
as shown in experiments \cite{nature}.

The stable combination of two localized modes $\psi_{\pm}$ will be determined by the quartic order free energy
\be
f_4=\beta_1(|\psi_{+}|^4+|\psi_{-}|^4)+\beta_2|\psi_{+}|^2|\psi_{-}|^2
\ee
with coefficients $\beta_{1,2}$, and $\beta_1>0$, $\beta_2>-2\beta_1$ for the stability of the free energy (see Appendix). When $\beta_2>2\beta_1$ the stable state will break the inversion symmetry and choose one of the localized modes of $\psi_{\pm}$. When $\beta_2<2\beta_1$ the stable state will preserve the inversion symmetry to form a symmetric combination of $\psi_{\pm}$, while the in-plane translation symmetry along $\bm q\propto\bm B\times\hat{\bm z}$ direction is spontaneously broken. In the decoupling limit $\mathcal{J}\to 0$, the equilibrium state will preserve the inversion symmetry to compensate the orbital effect, hence we expect $\beta_2<2\beta_1$ generally holds when the Josephson coupling is not too strong. 
The orbital FFLO state would be a Josephson vortex formed by localized modes on middle layers.

\textit{Conclusion}---
In this work, we generalize the orbital FFLO states theoretically derived in the bilayer superconductors \cite{CXL,GWQ} to multilayers. We find that due to the finite size effect, the middle layers become more favorable than other layers. Under an in-plane field, the multilayer Ising superconductor can be described by an effective bilayer model, with a field-driven phase transition from the conventional pairing state to the orbital FFLO state.
Our theory can be applied in transition metal dichalcogenide layers such as NbSe$_2$ \cite{nature,XXi,WYH}.

\textit{Acknowledgement}---The author thanks Puhua Wan, Jianting Ye and Chao-Xing Liu for inspiring discussions. The author thanks Puhua Wan and Jianting Ye for sharing the experimental data.
The author acknowledges the National Natural Science Foundation of China (Grant. No. 12174021) for the financial support.

\begin{widetext}
    \begin{appendix}
        \section{Model}
We consider the $N$-layer Lawrence-Doniach (LD) model with the following free energy density per area
\bea\label{eq_LD}
f=\sum_{l=1}^{N}\left\{\alpha|\psi_l|^2+\frac{\beta}{2}|\psi_l|^4+\frac{|(\nabla_{\parallel}-2ie\bm A_l)\psi_l|^2}{2m}\right\}
-J\sum_{l=1}^{N-1}(\psi_{l}^{*}\psi_{l+1}+\psi_{l+1}^{*}\psi_{l})\exp\left(2ie\int_{ld}^{(l+1)d} A_z dz\right),
\eea
and introduce the following three quantities $[\alpha=\alpha_0(T-T_{c1})]$
\be\label{eq_units}
T_0= \frac{J}{\alpha_0},\quad B_0=\frac{\Phi_0 q_0}{2\pi d},\quad q_0=\sqrt{2mJ}.
\ee

The Bogouliubov de Gennes Hamiltonian of a monolayer Ising superconductor is
\be
H=\xi\tau_z +\epsilon\Delta_{so}\sigma_z +\mu B\sigma_x\tau_z+\Delta \sigma_y\tau_y
\ee
where $\xi$ is the kinetic energy, $\epsilon=\pm$ is the valley index, $\Delta$ is the pairing amplitude, $\bm\sigma$ and $\bm\tau$ are Pauli matrices in spin and particle-hole spaces respectively. This Hamiltonian can have four eigenvalues $\{\pm E_{+},\pm E_{-}\}$ for each momentum, and the Bogouliubov spectrum is always fully gapped
\be
E_{\pm}=\sqrt{\xi^2+\Delta_{so}^2+(\mu B)^2+\Delta^2\pm 2\sqrt{(\xi\Delta_{so})^2+(\xi\mu B)^2+(\mu B\Delta)^2}}.
\ee
Thus the free energy per area reads $f_1=\Delta^2/g-T\sum_E\log(1+e^{-E/T})$ with interaction parameter $g$, and
\be
\alpha =\left.\frac{\partial^2 f_1}{\partial\Delta^2}\right|_{\Delta=0}=N_0\left\{
\log\left(\frac{T}{T_{c1}}\right)-\frac{B^2}{B^2+B_{so}^2}F\left(\frac{\mu\sqrt{B^2+B_{so}^2}}{\pi T}\right)
\right\}.
\ee
Here $T_{c1}=\omega_D\exp(-1/N_0g)$ is the monolayer critical temperature at zero field, with Debye frequency $\omega_D$ and monolayer density of states $N_0$. And $B_{so}\equiv\Delta_{so}/\mu$ is the Ising limit.
the special function is defined in terms of the digamma function $\Psi(x)$ as
\be
F(x)={\rm Re}\left\{\Psi\left(\frac{1}{2}\right)-\Psi\left[\frac{1}{2}(1+ix)\right]\right\}.
\ee

\section{Order parameter profiles}
At zero field, the linearized Ginzburg-Landau equation is the following eigenvalue problem
\bea
J\psi_2=\alpha\psi_1,\quad
J\psi_{N-1}=\alpha\psi_{N},\quad
J(\psi_{l-1}+\psi_{l+1})=\alpha\psi_l.\quad (l=2,\dots,N-1)
\eea
The order parameter can be worked out
\be\label{eq_psi0}
\psi_l=\Delta\sin\left(\frac{\pi l}{N+1}\right).
\ee
In order to see this, we notice the following trigonometric identities
\bea
\psi_2=2\cos\left(\frac{\pi}{N+1}\right)\psi_1,\quad
\psi_{N-1}=2\cos\left(\frac{\pi}{N+1}\right)\psi_{N},\\
\psi_{l-1}+\psi_{l+1}=2\cos\left(\frac{\pi}{N+1}\right)\psi_l.\quad (l=2,\dots,N-1)
\eea
Since $\alpha=\alpha_0(T-T_{c1})$, then
at zero field, superconductivity occurs at temperatures below the critical temperature
\be\label{eq_TcN}
T_{cN}=T_{c1}+2T_0\cos\left(\frac{\pi}{N+1}\right).
\ee

At weak in-plane fields, we treat $B$ as a perturbation and calculate the free energy up to $\Delta^2$ as
\bea
f=\frac{\Delta^2}{2}\left\{\alpha -2J\cos\left(\frac{\pi}{N+1}\right)+J\frac{B^2}{B_0^2}c_N\right\},
\eea
where the dimensionless coefficient is
\be\label{eq_cn}
c_N=
-\frac{1}{24} \left[\left(N^2+2 N+3\right) \cos \left(\frac{2 \pi }{N+1}\right)-N^2-2 N+9\right] \csc ^2\left(\frac{\pi }{N+1}\right).
\ee
Hence the critical temperature at weak in-plane fields reads
\bea\label{eq_tcnb}
T_{cN}(B)=T_{cN}-c_N T_0(B/B_0)^2.
\eea
Values of $c_N$ are calculated and listed in Table. \ref{table0}, and have been used in the fitting of Fig. 1a of the maintext.

Under in-plane fields, the linearized Ginzburg-Landau equation is the eigenvalue problem of difference equation
\bea
-\frac{|\bm q-2e\bm A_l|^2}{2m}\psi_l+J(\psi_{l-1}+\psi_{l+1})=\alpha\psi_l,\quad
\bm A_l=[l-\frac{1}{2}(N+1)]d\bm B\times\hat{\bm z}.\quad (l=2,\dots,N-1)
\eea
When the field is strong enough (intermediate), we can use continuum approximation
\bea
\psi_{l-1}+\psi_{l+1}-2\psi_l\approx \frac{d^2}{dl^2}\psi_l,
\eea
and the difference equation becomes a differential equation
\bea
-\frac{|l-l_0|^2}{2m}\left(\frac{2\pi Bd}{\Phi_0}\right)^2\psi_l+J\frac{d^2}{dl^2}\psi_l=(\alpha-2J)\psi_l,\quad
l_0=\frac{1}{2}(N+1)+\frac{\Phi_0 q}{2\pi Bd}.
\eea
The solution will be Gaussian
\be
\psi_l=\Delta \exp\left(-\frac{1}{2}\frac{B}{B_0}|l-l_0|^2\right).
\ee
When $q=-\pi Bd/\Phi$, the order parameter localizes on layer $l_0=\frac{1}{2} N$, and when $q=+\pi Bd/\Phi$, the order parameter localizes on layer $l_0=\frac{1}{2} N+1$.
The width of the order parameter profile is $l_B=\sqrt{B_0/B}$.

For comparison, we can also estimate the width of the order parameter profile Eq. (\ref{eq_psi0}) at zero field. To do this, we notice that the peak of Eq. (\ref{eq_psi0}) is at $l_c=\frac{1}{2}(N+1)$, which is a half integer if $N$ is even and an integer if $N$ odd. 
We thus expand Eq. (\ref{eq_psi0}) around $l=l_c$ as
\be
\frac{\psi_{\delta l+l_c}}{\Delta}=\cos\left(\frac{\pi \delta l}{N+1}\right)=1-\frac{1}{2}\left(\frac{\pi \delta l}{N+1}\right)^2+O(\delta l^4)
\ee
Hence the peak width can be argued as
\be
\left(\frac{\pi \delta l}{N+1}\right)^2\sim\frac{1}{N+1}\Rightarrow
\delta l\sim \frac{\sqrt{N+1}}{\pi}.
\ee

\section{Effective bilayer model}
The point group of the bilayer and bulk transition metal dichalcogenide is $D_{3d}$ with symmetry generators: Threefold in-plane rotation $C_{3z}$, vertical mirror $M_x$ and twofold out-of-plane rotation $C_{2x}$.
Symmetry operations on the two localized modes $\psi_{\pm}$, Cooper pair momentum $\bm q=(q_x,q_y)$ and in-plane magnetic field $\bm B=(B_x,B_y)$ are 
\bea
C_{3z}:\ \psi_{\pm}\to\psi_{\pm},\ (q_x,q_y)\to \left(-\frac{1}{2}q_x-\frac{\sqrt{3}}{2}q_y,\frac{\sqrt{3}}{2}q_x-\frac{1}{2}q_y\right),\\\nonumber
(B_x,B_y)\to \left(-\frac{1}{2}B_x-\frac{\sqrt{3}}{2}B_y,\frac{\sqrt{3}}{2}B_x-\frac{1}{2}B_y\right),\\\nonumber
M_x:\ \psi_{\pm}\to\psi_{\pm},\ (q_x,q_y)\to (-q_x,q_y),\ (B_x,B_y)\to (B_x,-B_y),\\\nonumber
C_{2x}:\ \psi_{\pm}\to\psi_{\mp},\ (q_x,q_y)\to (q_x,-q_y),\ (B_x,B_y)\to (B_x,-B_y),\\\nonumber
\mathcal{T}:\ \psi_{\pm}\to\psi_{\pm}^{*},\ (q_x,q_y)\to -(q_x,q_y),\ (B_x,B_y)\to -(B_x,B_y).
\eea
The inversion can be obtained by group multiplication $I=M_xC_{2x}: \psi_{+}\leftrightarrow\psi_{-},\bm q\to -\bm q,\bm B\to\bm B$.

The free energy is invariant under the point group $D_{3d}$, so is the Hessian matrix, namely $U(g)\mathcal{H}(g\bm q,\overline{g}\bm B)U^{-1}(g)=\mathcal{H}(\bm q,\bm B)$ with $\overline{g}=\det(g)g$ and $U(C_{2x})=\tau_x,U(C_{3z})=U(M_x)=\tau_0$.
In order to calculate the invariants, we introduce $q_{\pm}=q_x\pm iq_y$, $B_{\pm}=B_x\pm iB_y$, and $\omega=\exp(2\pi i/3)$, then
\bea
C_{3z}:\ \psi_{\pm}\to\psi_{\pm},\ q_{\pm}\to \omega^{\pm} q_{\pm},\ B_{\pm}\to \omega^{\pm}B_{\pm},\\\nonumber
M_x:\ \psi_{\pm}\to\psi_{\pm},\ q_{\pm}\to -q_{\mp},\ B_{\pm}\to B_{\mp},\\\nonumber
C_{2x}:\ \psi_{\pm}\to\psi_{\mp},\ q_{\pm}\to q_{\mp},\ B_{\pm}\to B_{\mp},\\\nonumber
\mathcal{T}:\ \psi_{\pm}\to\psi_{\pm}^{*},\ \bm q\to -\bm q,\ \bm B\to -\bm B.
\eea
We thus obtain the following invariants for example
\be
\tau_0,\ q^2,\ B^2,\ {\rm Re}(q_{+}^2B^4_{+}),\ {\rm Re}(q_{+}^4B^2_{+}),\ 
\tau_x,\ {\rm Im}(q_{+}B^2_{+})\tau_y,\ {\rm Im}(q_{+}B_{-})\tau_z,\ {\rm Im}(q_{+}^3B^3_{+})\tau_z,\dots
\ee
In general, the invariants would have the following forms ($n,k=0,1,2,\dots$)
\be
q^{2n}B^{2k},\ {\rm Re}(q_{+}^{2n}B^{6k-2n}_{+}),\ 
q^{2n}B^{2k}\tau_x,\ {\rm Im}(q_{+}^{2n-1}B^{6k-2n-2}_{+})\tau_y,\ {\rm Im}(q_{+}^{2n-1}B_{-}^{2n-1-6k})\tau_z,\ {\rm Im}(q_{+}^{2n-1}B^{6k+1-2n}_{+})\tau_z,\dots
\ee

Hence we arrive at the Hessian matrix $\mathcal{H}=\mathcal{H}_0+\mathcal{H}_1$ as shown in the maintext:
\bea
    \mathcal{H}_0=a+b(\hat{\bm z}\times\bm B)\cdot\bm q\tau_z+cq^2-\mathcal{J}\tau_x,\
    a=r(T-T_a)+\chi B^2,\\
    \mathcal{H}_1=\lambda_1{\rm Re}(q_{+}^2B_{+}^4)+\lambda_2{\rm Re}(q_{+}^4B_{+}^2)+\lambda_3{\rm Im}(q_{+}B^2_{+})\tau_y.
\eea
Hence, we can derive the field-dependent optimal Cooper pair momentum and critical temperature
\bea\label{eq_q}
\bm q=\frac{b}{2c}(\bm B\times\hat{\bm z}){\rm Re}\sqrt{1-\left(\frac{B^{*}}{B}\right)^4},\\
T_c(B,\varphi)=T_c(B)+\Delta T_c(B,\varphi),
\eea
with $B^*={\sqrt{2c|\mathcal{J}|}}/{b},$ $\bm B=B(\cos\varphi,\sin\varphi)$. The isotropic part of the critical temperature is
\bea\label{eq_tcb}
T_c(B)=T_a-\frac{\chi}{r}B^2+\frac{b^2}{4cr}\times
\begin{cases}
    2|B^*|^2 & B<B^*\\
    \left(B^2+{|B^*|^4}/{B^2}\right) & B>B^*
\end{cases}
\eea
And the anisotropic part is sixfold
\bea
\Delta T_c(B,\varphi)=\lambda(B)\theta(B-B^*)\cos(6\varphi),
\eea
with
\bea
\lambda(B)=\frac{1}{r}(\lambda_1\tau^2-\lambda_2\tau^4)B^6+\frac{1}{r}\frac{\lambda_3^2 B^4}{\sqrt{b^4/c^2+2\lambda_3^2\tau^2B^2}},\
\tau=\frac{b}{2c}\sqrt{1-\left(\frac{B^{*}}{B}\right)^4}.
\eea

To fit the numerical data in Fig. 1 of the maintext, we rewrite Eq. (\ref{eq_q}) as the dimensionless form
\be\label{eq_q1}
\frac{\bm q}{q_0}=\rho\frac{\bm B\times\hat{\bm z}}{B_0}{\rm Re}\sqrt{1-\left(\frac{B^{*}}{B}\right)^4}
\ee
with the dimensionless parameters $\rho=B_0b/(2cq_0)$ and $B^*/B_0$ listed in Table. \ref{table0} for different $N$.

To fit the experimental data in Fig. 3 of the maintext, we rewrite Eq. (\ref{eq_tcb}) as the dimensionless form
\be\label{eq_tcb1}
\frac{T_c(B)}{T_c}=1-\gamma_1\left(\frac{B}{B^*}\right)^2+\gamma_2\left[\left(\frac{B}{B^*}\right)^2+\left(\frac{B^*}{B}\right)^2-2\right]\theta(B-B^*),\quad
T_c=T_a+\mathcal{J}/r,
\ee
with two dimensionless parameters $\gamma_{1}=\chi |B^*|^2/r$, $\gamma_2=\mathcal{J}/(2r)$ and one parameter $B^*$ of the field dimension. The optimal fitting gives rise to
\be
\gamma_1=0.1235,\ \gamma_2=0.0882,\ B^*=4.0911{\rm T}.
\ee

\begin{table*}
\centering
\caption{Fitting parameters of Eqs. (\ref{eq_cn}) and (\ref{eq_q1}) for numerical data in Fig. \ref{fig1} of the maintext.}
\begin{center}  
\begin{tabular}{c||c|c|c|c|c|c|c|c|c|c}  
\hline  
$N$   & 2 & 4&6&8&10&12&14&16&18&20  \\ \hline   
$c_N$ & $1/4$& 0.8028& 1.5940& 2.6424& 3.9507& 5.5197& 7.3499& 9.4413& 11.7939& 14.4079 \\ \hline 
$\rho$ & $1/2$ &0.5640&0.6280&0.7020&0.7020&0.8000&0.8000&0.8500&0.8500&0.8500 \\ \hline 
$B^*_N/B_0$ &  $\sqrt{2}$&1.0350&0.7839& 0.6231& 0.5126& 0.4422& 0.3819& 0.3417& 0.3015& 0.2714\\ \hline 
\end{tabular}  
\end{center} 
\label{table0}
\end{table*}

Next we consider the quartic free energy of the effective bilayer model
\be
f_4=\beta_1(|\psi_{+}|^4+|\psi_{-}|^4)+\beta_2|\psi_{+}|^2|\psi_{-}|^2.
\ee
We introduce the polar coordinate $\psi_{+}=\psi\cos\theta$ and $\psi_{-}=\psi\sin\theta$ to rewrite the free energy above
\be
f_4=\beta_1|\psi|^4+\frac{1}{4}(\beta_2-2\beta_1)|\psi|^4\sin^2(2\theta).
\ee
For stability, we require the coefficient of $|\psi|^4$ term to stay positive for various $\theta$. When $\theta=0$ we find $\beta_1>0$. When $\sin^2(2\theta)=1$, we find
\be
\beta_1+\frac{1}{4}(\beta_2-2\beta_1)>0\Rightarrow\beta_2>-2\beta_1.
\ee
    \end{appendix}
\end{widetext}

\end{document}